\documentclass[twocolumn,aps,prx,showpacs,bibfootnote,preprintnumbers, mathtools, superscriptaddress, amsmath,amssymb]{revtex4-1}

\usepackage{amsmath,amsfonts,amsthm,bm}
\usepackage[caption=false]{subfig}
\usepackage{graphicx}
\usepackage{xcolor}

\begin{document}

\author{Gerard McCaul}
\email{gmccaul@tulane.edu}
\affiliation{Tulane University, New Orleans, LA 70118, USA}

\author{Kurt Jacobs} 
\affiliation{U.S. Army Research Laboratory, Sensors and Electron Devices Directorate, Adelphi, Maryland 20783, USA} 
\affiliation{Department of Physics, University of Massachusetts at Boston, Boston, MA 02125, USA} 

\author{Denys I. Bondar}
\email{dbondar@tulane.edu}
\affiliation{Tulane University, New Orleans, LA 70118, USA}

\title{Beating the House: Fast Simulation of Dissipative Quantum Systems with Ensemble Rank Truncation}

\begin{abstract}
  We introduce a new technique for the simulation of dissipative quantum systems. This method is composed of an approximate decomposition of the Lindblad equation into a Kraus map, from which one can define an ensemble of wavefunctions. Using principal component analysis, this ensemble can be truncated to a manageable size without sacrificing numerical accuracy. We term this method \emph{Ensemble Rank Truncation} (ERT), and find that in the regime of weak coupling, this method is able to outperform existing wavefunction Monte-Carlo methods by an order of magnitude in both accuracy and speed. We also explore the possibility of combining ERT with approximate techniques for simulating large systems (such as Matrix Product States (MPS)), and show that in many cases this approach will be more efficient than directly expressing the density matrix in its MPS form. We expect the ERT technique to be of practical interest  when simulating dissipative systems for quantum information, metrology and thermodynamics.
\end{abstract}

\date{\today}

\maketitle

\section{Introduction}

 In the emerging field of quantum technologies, environments play a crucial role. Their most obvious contribution is to introduce noise which destroys the delicate coherences necessary for truly quantum effects, but system-environment interactions can also harbour some surprising benefits when carefully controlled \cite{PhysRevA.98.042133,PhysRevLett.120.230404}. For example, one may use dissipation to prepare entangled states ~\cite{PhysRevLett.106.090502,PhysRevLett.117.040501}, induce nonreciprocal photon transmission ~\cite{PhysRevX.10.021038} and amplification ~\cite{PhysRevX.5.021025},  exploit cooperative effects to \emph{reduce} entropy ~\cite{PhysRevLett.107.130404}, as well as {engineer dynamics  \cite{vuglar_nonconservative_2018}}. Given the manifold challenges and opportunities presented by open quantum systems, it is of vital importance to have a fast and scalable method for simulating them. 
 
The first problem one encounters in modelling open systems is how to account for the fact that a full system +environment amalgam exists in a Hilbert space far too large to be simulated exactly. One must instead focus on an effective description of the system of interest, eliminating the environmental degrees of freedom. A number of approaches have been developed to tackle this problem, with one of the earliest examples being the Redfield equation ~\cite{REDFIELD19651,Redfield:1957:TRP:1661805.1661808,redfield2}. While this equation provides a good approximation to dynamics at weak coupling, it does not guarantee the positivity of the density matrix ~\cite{Redfieldbook}, and as such { has remained controversial~\cite{McCauley2020}.}

Another fruitful avenue has been to approach the problem of environmental coupling through the path-integral formalism of the Feynman-Vernon influence functional ~\cite{Feynman-Vernon-1963,InfluenceFunctionalarxiv}. This technique has produced a great number of important  results ~\cite{Kleinertbook,KLEINERT1995, Tsusaka1999,Smith1987,Makri1989,Allinger1989, Bhadra2016,McDowell2000}, including the derivation of a number of distinct dynamical equations, such as the quantum Langevin ~\cite{Caldeira1983,Ford-Kac-JST-1987,Gardiner-1988,Sebastian1981,Leggett1987,van_Kampen-1997},  stochastic Schr\"{o}dinger ~\cite{Orth2013,Orth2010}, quantum Smoluchowski ~\cite{Ankerhold2001,Maier2010} and stochastic Liouville-von Neumann ~\cite{Stockburger2002, Stockburger2004, Stockburger2017, PhysRevB.95.125124,PhysRevB.97.224310} equations. 

While influence functional techniques (being non-perturbative) are particularly useful at strong environmental coupling, in the weak coupling regime it is far more common to employ the Lindblad master equation ~~\cite{Jacobs14, Gardiner10, Petruccione-open-systems-book}. This equation represents the most general description for a Markovian process that preserves the essential features of the density matrix (complete positivity and trace preservation) ~\cite{Lindblad1976,Gorini}. It is this generality that has led to the Lindblad equation acquiring its status as the `default' approach to simulating open systems ~\cite{Monzanoreview}, and has found application in a very broad range of physical settings \cite{PhysRevA.97.053841, PhysRevLett.81.2594,RobustEntanglement,environmentassistedquantumwalks, Plenio_2008,PhysRevA.78.042307}. Indeed, it has recently been shown that the secular approximation \cite{scali2020local} used to derive the Lindblad equation is unnecessary, giving the equation a much broader regime of applicability than was previously supposed ~\cite{McCauley2020, Nathan20}. { Recent work has also shown that the Lindblad equation can be used for systems undergoing various kinds of broadband control~\cite{PhysRevResearch.2.013049}.}

Given all this, an efficient scheme for computing Lindblad dynamics is {of great importance across the many disciplines that use it, enabling the simulation of larger systems which may in turn lead to new discoveries. Achieving this however is problematic, as simulating this master equation requires the density matrix, which for a Hilbert space of size $N_H$ has $N_H^2$ elements. This means that simulating the Lindblad equation requires $O(N^3_H)$ operations compared to the $O(N^2_H)$ operations required to evolve the  Schr{\"o}dinger equation.} 

Given the exponential increase in $N_H$ that occurs as one adds particles to a many-body system, the difference between $O(N^3_H)$ and  $O(N^2_H)$  translates into a tremendous disparity in simulation runtime between the Lindblad and Schr{\"o}dinger equations for even a moderately large system. While in the last decade enormous strides have been made in the approximate simulation of large systems (see for example recent work simulating strongly interacting quantum thermal machines \cite{PhysRevX.10.031040}), the methods employed do not address the fundamental difference in scaling between density matrices and wavefunctions. Instead, the density matrix is flattened into a vector \cite{PhysRevLett.93.207204, zwolak_mixed-state_2004, luchnikov_simulation_2019} and the power of tensor network decompositions \cite{SCHOLLWOCK201196} are leveraged to efficiently evolve this large system. In this sense, one has simply transposed the density matrix into a wavefunction living in a Hilbert space of size $N^2_H$ rather than $N_H$. For this reason, there is a powerful motivation to express the Lindblad equation using wavefunctions of size $N_H$ rather than the full density matrix.

The standard solution to this issue has long been to employ a Wavefunction Monte-Carlo (WMC) method (although similar stochastic methods using classical noise have been proposed \cite{PhysRevLett.118.140403}). This  technique exploits the fact that the evolution of the Lindblad equation is equivalent to averaging over an ensemble of stochastic wavefunction evolutions ~\cite{Srinivas81, Gisin84, Diosi86, Belavkin87, Carmichael89, Wiseman93}. If the number of realisations $n_{\rm traj}$ needed to converge the averaging to an accurate answer is much less than $N_H$, then WMC provides an efficient alternative to the exact evolution of the Lindblad equation ~\cite{Hegerfeldt92, Molmer93}. Of course, like all stochastic methods the number of trajectories needed to converge a WMC result will depend on the specifics of the model system, as well as the length of time being simulated. Consequently, there will be circumstances in which the $n_{\rm traj}$ required will match or even exceed $N_H$, negating the advantages of employing WMC. 

In this paper we introduce a new wavefunction based method for describing Lindblad dynamics, which we term \emph{Ensemble Rank Truncation} (ERT). This method is based on deriving a general expression for the Kraus map of an infinitesimal evolution of the Lindblad equation {(up to the second order error in the time step, $O({\rm d}t^2)$)}. This mapping provides a deterministic procedure for constructing an ensemble of wavefunctions which together accurately describe system expectations. The size of the ensemble generated through this method increases exponentially with time, necessitating a second procedure to counteract this. Once the size of the ensemble has exceeded a prespecified rank $R$, principal component analysis is performed to generate a truncated set of wavefunctions in the basis which most efficiently represents the density matrix. This method differs sharply from WMC in two important respects- first, its dynamics are entirely deterministic and do not require any stochastic averaging. This leads to the second distinguishing property, namely that the size of the ensemble produced in this method tends to be much smaller than that required by WMC, with $R \ll n_{\rm traj}$. {Our results show that these features allow ERT to `beat the house', providing a great improvement over WMC for systems with weak environmental noise (damping and/or dephasing).}

{A full derivation of our ERT method is presented in Sec.~\ref{sec:Derivation}, together with a formal demonstration that using ERT with Matrix Product State (MPS) simulations of pure states is more efficient than directly using an MPS representation for a mixed-state evolution.} Section~\ref{sec:results} demonstrates the efficacy of ERT in three separate systems -- a Heisenberg spin chain, a set of atoms in a cavity, and a dissipative Fermi-Hubbard model. In particular, we show here that ERT can provide order of magnitude improvements in both speed and accuracy \emph{simultaneously} as compared to the standard WMF approach. We conclude in Sec.~\ref{sec:discussion} with a discussion of the results presented here, and suggest potential improvements and extensions to the ERT method.

\section{Ensemble Rank Truncation \label{sec:Derivation}}
In this section, we derive the approximation we term \emph{Ensemble Rank Truncation} (ERT). This approximation consists of two parts -- an approximate Kraus map which allows the dissipative evolution of the system to be represented with an ensemble of wavefunctions, and a truncation of that ensemble to its principal components, which is equivalent to excluding components of the density matrix above a certain rank. 

\subsection{Approximate Kraus operators}
It is well known that the time evolution of any system (closed or open) must be a Completely Positive Trace Preserving (CPTP) map~~\cite{wilde2013quantum}, and that such maps on the density matrix can always be decomposed into a Kraus form~~\cite{1983}, given by:

\begin{equation}
    \hat{\rho}(t)=\sum^K_{k=1} \hat{M}_k\hat{\rho}(0)\hat{M}^\dagger_k
\end{equation}
where $\hat{\rho}(t)$ is the density matrix and the $\hat{M}_k$ are Kraus operators which must collectively satisfy 
\begin{equation}
\label{eq:Krausnormalisation}
\sum^K_{k=1} \hat{M}^\dagger_k\hat{M}_k =\mathbb{I}.   
\end{equation}

Expressing a dissipative evolution in this way has obvious advantages compared to the usual master equation, for which the Kraus form may be formally regarded as a solution. Furthermore, if one begins with a pure state $\hat{\rho}(0)= |\psi\rangle\langle \psi|$, it is possible to express the expectation of an operator $\hat{O}$ purely in terms of wavefunctions, rather than the full density matrix. To do so, one defines a set of wavefunctions $\left\{|\psi_k\rangle\right\}=\{\hat{M}_k|\psi\rangle\}_{k=1}^K$, with which expectations can be expressed as:
\begin{equation}
\label{eq:expectationswithpsi}
\langle \hat{O}\rangle =\sum^K_{k=1} \langle{\psi_k}|\hat{O}|{\psi_k}\rangle.    
\end{equation}
The advantage of this form is that rather than an $N_H\times N_H$ density matrix, one instead requires only $K$ wavefunctions of dimension $N_H$. In the { typical} case $N_H \gg K$, it is far more efficient to calculate expectations via Eq.~\eqref{eq:expectationswithpsi} than with the full density matrix. Of course, this approach is only possible if one knows the $\hat{M}_k$ that describe the system of interest. 

Unfortunately, while there are many existence proofs for Kraus operators describing open system evolutions~~\cite{alicki1987quantum,Tong2006,Arsenijevic2018} (including for controllability of those systems~~\cite{Wu_2007}), the number and explicit form of these operators are known only for a few special cases~~\cite{nielsen2010quantum}. This is a problem, as one often wishes to solve for the dynamics of a system described by a Lindblad master equation~~\cite{Petruccione-open-systems-book}, given by
\begin{align}\label{eq:LindbladMasterEq}
	\frac{d\hat{\rho}(t)}{dt} &= -\frac{i}{\hbar} [\hat{H}, \hat{\rho}(t)] 
		+ \frac{1}{2}\sum_{k=1}^K \left( 2 \hat{A}_k \hat{\rho}(t) \hat{A}_k^{\dagger} 
			-\left\{\hat{\rho}(t), \hat{A}_k^{\dagger} \hat{A}_k\right\} \right)
\end{align}
where $\hat{H}$ is the Hamiltonian and the $\hat{A}_k$ are the environmental dissipators. As discussed previously, for large Hilbert space dimension $N_H$, the calculation of the density matrix as compared to  a set of $K$ wavefunctions is extremely expensive. We would therefore like to find a way to express this equation in a Kraus form. 

To do so, we first discretise Eq. \eqref{eq:LindbladMasterEq}:
\begin{align}\label{eq:Lindbladonestep}
	\hat{\rho}(t + {\rm d }t)&= \hat{\rho}(t) -{\rm d}t \frac{i}{\hbar} [\hat{H}, \hat{\rho}(t)]+ O\left({\rm d}t^2 \right) \notag\\
		&+ \frac{{\rm d}t}{2}\sum_{k=1}^K \left( 2\hat{A}_k \hat{\rho}(t) \hat{A}_k^{\dagger} 
			- \left\{\hat{\rho}(t), \hat{A}_k^{\dagger} \hat{A}_k\right\}  \right).
\end{align}
It is then possible to represent this expression as a Kraus map:
\begin{align}\label{eq:mainformula}
    \hat{\rho}(t + {\rm d }t) =& \frac{1}{2K} \sum_{k=1}^K \left(
            \hat{\mathcal{U}}_k \hat{\rho}(t) \hat{\mathcal{U}}_k^{\dagger}
            +
            \hat{\mathcal{V}}_k \hat{\rho}(t) \hat{\mathcal{V}}_k^{\dagger}
        \right)  + O\left({\rm d}t^2 \right),
\end{align}
where the operators $\hat{\mathcal{U}}_k$ and $\hat{\mathcal{V}}_k$ are defined as
\begin{align}
\hat{\mathcal{U}}_k &={\rm e}^{{\rm d}t \hat{J}_k-i\sqrt{K{\rm d}t}\hat{A}_k} \\
\hat{\mathcal{V}}_k &={\rm e}^{{\rm d}t \hat{J}_k+i\sqrt{K{\rm d}t}\hat{A}_k} \\ 
\hat{J}_k&=-\frac{i}{\hbar}\hat{H}+\frac{K}{2}(\hat{A}^2_k-\hat{A}_k^\dagger \hat{A}_k).
	\end{align}
Note that in the case of Hermitian dissipators $\hat{A}^\dagger_k=\hat{A}_k$, the Kraus operators are unitary, and the infinitesimal evolution has the form of a random unitary channel~\cite{Audenaert2008}.

In order to prove Eq.~\eqref{eq:mainformula}, we use the Zassenhaus~~\cite{Casas2012} approximation,
\begin{equation}
{\rm{e}}^{\Delta(X+Y)}={\rm{e}}^{\Delta X}{\rm e}^{\Delta  Y}{\rm{e}}^{-\frac{\Delta^2}{2}[X,Y] + O(\Delta^3)},
\end{equation}
and after substituting $\Delta=\sqrt{{\rm d}t}$, $X=\hat{J}_k\sqrt{{\rm d}t}$ and $Y=\mp i\sqrt{K}\hat{A}_k$, we obtain the following representations for the operators to $O\left(dt^2 \right)$:
\begin{align}
    \hat{\mathcal{U}}_k &=e^{\hat{J}_k {\rm d}t}U^-_k+O\left(dt^2 \right), \quad  \hat{\mathcal{V}}_k =e^{\hat{J}_k {\rm d}t}U^+_k +O\left(dt^2 \right)\\
    \bar{U}^\pm_k &= e^{\pm i\hat{A}_k \sqrt{Kdt}} e^{\pm{\rm d}t^{\frac{3}{2}}\frac{\sqrt{K}}{\hbar}[\hat{J}_k,\hat{A}_k]} \label{eq:Uplusminus}.
\end{align}
These approximations can be used to expand a single term in the sum on the RHS of Eq.~\eqref{eq:mainformula} to  $O({\rm d}t^2)$ :
\begin{equation}
\label{eq:singleterm}
     \hat{\mathcal{U}}_k \hat{\rho}(t) \hat{\mathcal{U}}_k^{\dagger} + \hat{\mathcal{V}}_k \hat{\rho}(t) \hat{\mathcal{V}}_k^{\dagger}={\rm e}^{{\rm d}t \hat{J}_k}\hat{\Gamma}_k(t){\rm e}^{{\rm d}t \hat{J}_k^\dagger } +O({\rm d}t^2),
\end{equation}
and after some tedious algebra, one finds
\begin{align}
    \hat{\Gamma}_k(t)&=\hat{\rho}(t) + K{\rm d }t\left(\hat{A}_k\hat{\rho}(t) \hat{A}_k^\dagger -\hat{A}^2_k\hat{\rho}(t)\right) 
    +{\rm  h.c}
\label{eq:Xexpansion}.
\end{align}
We now perform a final expansion of ${\rm e}^{{\rm d}t \hat{J}_k}$ to  $O({\rm d}t^2)$ and substitute both this and $ \hat{\Gamma}_k(t)$ back into Eq.~\eqref{eq:singleterm} to obtain:
\begin{align}
    \hat{\mathcal{U}}_k & \hat{\rho}(t) \hat{\mathcal{U}}_k^{\dagger} + \hat{\mathcal{V}}_k \hat{\rho}(t) \hat{\mathcal{V}}_k^{\dagger}= 2\hat{\rho}(t)- 2{\rm d}t \frac{i}{\hbar} [\hat{H}, \hat{\rho}(t)] \notag \\ &+ K{\rm d }t\left(2\hat{A}_k \hat{\rho}(t) \hat{A}_k^{\dagger} 
			- \left\{\hat{\rho}(t),\hat{A}_k^{\dagger} \hat{A}_k\right\} \right) +O({\rm d}t^2).
\end{align}
Finally, inserting this equality into Eq.~\eqref{eq:mainformula} recovers Eq.~\eqref{eq:Lindbladonestep} to $O({\rm d}t^2)$, proving that $\hat{\mathcal{U}}_k$ and $\hat{\mathcal{V}}_k$ are the sought-for approximate Kraus operators for an infinitesimal evolution. Note also that these operators approximately satisfy Eq.~\eqref{eq:Krausnormalisation} with:
\begin{equation}
     \frac{1}{2K} \sum_{k=1}^K \left(
            \hat{\mathcal{U}}^{\dagger}_k \hat{\mathcal{U}}_k
            +
            \hat{\mathcal{V}}^{\dagger}_k  \hat{\mathcal{V}}_k
        \right)=\mathbb{I}+O({\rm d}t^2). 
\end{equation}

\subsection{Truncating the Ensemble} 
For a single time-step, the computational complexity of performing the matrix-matrix calculation for the exact Lindblad equation is $O(N_H^3)$, but employing the approximate Kraus form, obtaining the set of wavefunctions $\{|\psi_k\rangle\}=\{\hat{\mathcal{U}}_k|\psi \rangle,\hat{\mathcal{V}}_k|\psi \rangle \}_{k=1}^{K}$ needed to calculate expectations has a computational complexity of only $O(KN_H^2)$. Furthermore, using the approximate method one need only store $K N_H$ elements, whereas the density matrix requires $N_H^2$ elements. Clearly for a single step, if $N_H\gg K$, both the speed and storage requirements for the calculation will be improved by approximately a factor of $N_H$. Of course, the major problem with this method is that after each step, the number of wavefunctions increases by a factor of $2K$, so that after $M$ steps one must store a set of $(2K)^M$ wavefunctions. 

To overcome this bottleneck, we employ a further approximation, aimed at limiting the size of the wavefunction ensemble. First, let $R$ denote the pre-specified maximal rank of the density matrix $\hat{\rho}$. {The choice of $R$ is dictated by physics, the desired accuracy, and the available memory.}  For simplicity we assume that the initial density matrix is represented by
\begin{align}\label{EqIntialRhoMeth}
    \hat{\rho}(t) = |\psi\rangle\langle \psi|,
\end{align}
although the generalisation to a mixed density matrix is trivial. The procedure is then as follows -- we propagate according to Eq.~\eqref{eq:mainformula}, saving the set $\{|\psi_k\rangle\}$ at each step until there are $L=(2K)^M > R$ wavefunctions. Now the size of this set exceeds $R$, we orthogonalise them such that  
\begin{align}
    \hat{\rho}(t + M{\rm d}t) = \sum_{k=1}^{L} |\psi_k\rangle\langle \psi_k|= \sum_{k=1}^{L} |\bar{\psi}_k\rangle\langle \bar{\psi}_k|.
\end{align}
with $\{|\bar{\psi}_k\rangle\}$ being orthogonal but unnormalised. This orthogonalisation is achieved by
\begin{align}
    |\bar{\psi}_k\rangle = \sum_{l=1}^{L} U_{lk} |\psi_l\rangle,
\end{align}
where $U$ is a unitary matrix obtained from diagonalising the overlap matrix $S_{ij} = \langle \psi_i| \psi_j \rangle$, such that
\begin{align}
\label{eq:diageigenvalues}
    U^{\dagger} S U = \mathrm{diag}(w_1, \ldots, w_{L}).
\end{align}
Here we assume that the eigenvalues are arranged in descending order, $w_1 \geq \cdots \geq w_{L}$. Note that obtaining the overlap matrix involves $L^2$  wavefunction dot-products, meaning its computational cost is $O(L^2N_H)$. For $N_H \gg L^2$, this has a negligible cost relative to other operations. Furthermore, since the dimensionality of matrix $S$ is $L \times L$, its diagonalization is also computationally ``cheap'' and does not depend on the dimensionality $N_H$ of the Hilbert space.

\begin{figure}
\centering
\includegraphics[width=\columnwidth]{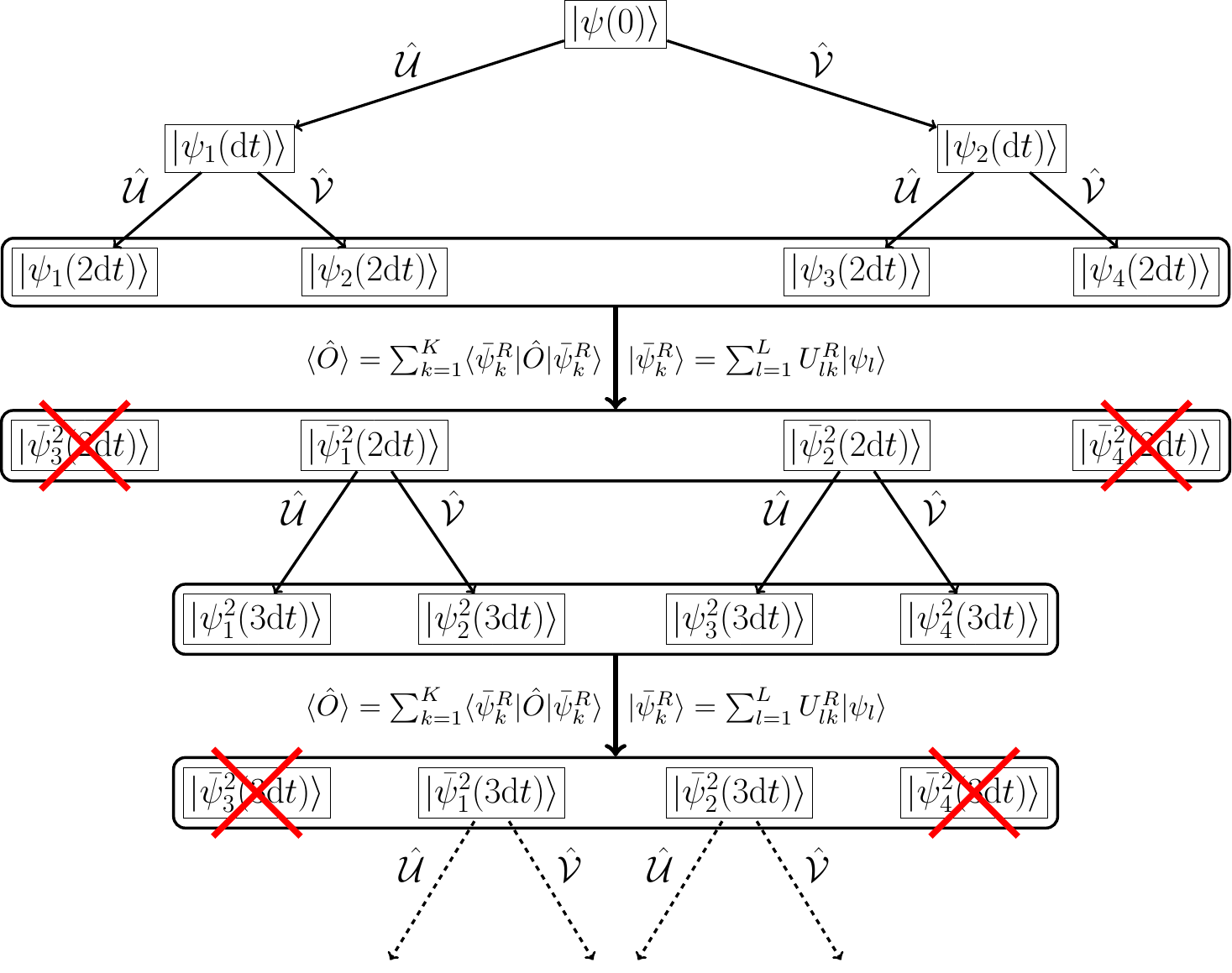}
\caption{Schematic showing the ERT process for a single dissipator and $R=2$. The initial wavefunction is propagated by the application of the $\mathcal{U}$ and $\mathcal{V}$ unitaries until the size of the set exceeds $R$.  Each time this occurs the orthogonalisation procedure is performed and the $R$ largest components of the set are retained for propagation to later steps.}
\label{fig:ERTexample}
\end{figure}

Having calculated $U$, we can truncate it to an $R\times L$ matrix $U^R$, using the eigenvectors associated with the $R$ largest eigenvalues. $U^R$ can then be used to generate an orthogonal ensemble of wavefunctions:
 \begin{align}
    |\bar{\psi}^R_k\rangle = \sum_{l=1}^{L} U^R_{lk} |\psi_l\rangle.
\end{align}
The generation of this truncated ensemble is akin to principal component analysis, a well-known statistical technique where $R$ is the number of principal axes \cite{jolliffe2002principal}. Using this procedure we need only store $\{|\bar{\psi}^R_k\rangle\}_{k=1}^R$, and after appropriate normalisation of this set we obtain the sought for ERT approximation,
\begin{align}
      \hat{\rho}(t + M{\rm d}t) 
    \approx \sum_{k=1}^{R} |\bar{\psi}^R_k\rangle\langle \bar{\psi}^R_k|.
\end{align}
Propagating to later steps then simply repeats the same process of orthogonalisation followed by truncation to $R$ wavefunctions, as shown in Fig.~\ref{fig:ERTexample}. Finally, we emphasize that stepping forward in time by generating the set $\{\hat{\mathcal{U}}_m|\bar{\psi}^R_{k} \rangle,\hat{\mathcal{V}}_{m} |\bar{\psi}^R_{k} \rangle  \}_{m=1}^{K}$ is a fully parellelisable calculation, so that if $2KR$ does not exceed the number of threads available, the time required for stepping forward will be approximately independent of the rank. { Furthermore, in the case of large systems where the calculation of matrix exponents is prohibitively costly, the effect of applying any given Kraus operator to a member of the ensemble can instead be characterised as an ODE }{ (with an explicit time-step of $\sqrt{{\rm d}t}$). A repeated propagation via Eq.~\eqref{eq:mainformula} can also be represented as a non-commutative Newton binomial \cite{wyss2017two, hosseini2019non}. This insight offers a potential route to speed up calculations.}

\subsection{Combining Ensemble Rank-Truncation with Matrix Product States}
It is worth taking a moment to place the ERT method in its proper context. It is the product of two separate approximations -- an approximation for the infinitesimal Kraus operators which allows one to characterise the dynamics with a set of wavefunctions, and a truncation of that set to its principal components at each timestep. The former approximation is to the best of our knowledge a novel representation of the density matrix evolution, while the reduction of the set of wavefunctions to their principal components is designed to constrain the size of the ensemble to a manageable size.

{It is interesting to note that both ERT and the matrix-product-state (MPS) representation ~\cite{vidal_efficient_2003, vidal_efficient_2004, ORUS2014117} make use of a singular-value decomposition to find the most efficient representation for a state. In the case of MPS it is used to find minimal-rank Schmidt decompositions to minimize the matrices that form the MPS representation for a pure state. Here it is used to minimize the number of pure states that are required to represent a mixed state.} 


{In fact, there is no reason why ERT cannot be performed on sets of wavefunctions evolved using time-dependent MPS (time evolving block decimation (TEBD)~\cite{vidal_efficient_2004}). 
One might ask whether combining ERT with an MPS representation for pure states provides any advantage versus directly using MPS to represent the evolving vectorized density matrix. To answer this, recall that in the MPS representation~\cite{vidal_efficient_2003, vidal_efficient_2004}, a quantum state $|\Psi\rangle$ is written as
\begin{align}\label{EqMPSWavefunc}
    |\Psi\rangle &= \sum_{i_1=0}^{d-1} \cdots \sum_{i_n=0}^{d-1} c_{i_1 \cdots i_n} |i_1\rangle \otimes\cdots\otimes |i_n\rangle, \notag\\
    c_{i_1 \cdots i_n} &= \sum_{\alpha_1, \ldots, \alpha_{n-1}=1}^{\chi}  \Gamma_{\alpha_1}^{[1]i_1}
    \lambda_{\alpha_1}^{[1]} \Gamma_{\alpha_1 \alpha_2}^{[2]i_2} \lambda_{\alpha_1}^{[2]}
    \cdots \Gamma_{\alpha_{n-1}}^{[n]i_n},
\end{align}
where $\{|i_1\rangle\otimes\cdots\otimes |i_n\rangle\}$ is a computational basis for an $n$-body  system, and $\chi$ is the Schmidt rank quantifying the degree of entanglement. In other words, a state $|\Psi\rangle$ is represented by $n$ tensors $\Gamma^{[1]}, \ldots, \Gamma^{[n]}$ (each with $\chi d$ elements) and $n-1$ length $\chi$ vectors $\lambda^{[1]}, \ldots, \lambda^{[n-1]}$.

We may apply the MPS represention directly to a density matrix $\hat{\rho}$ that has been flattened into a column vector $|\hat{\rho}\rangle_{\sharp}$~~\cite{zwolak_mixed-state_2004}. Using Eq.~\eqref{EqMPSWavefunc} the vectorised density matrix is expressed as:
\begin{align}\label{EqMPSDensity}
    |\hat{\rho}\rangle_{\sharp} &= \sum_{j_1=0}^{d^2-1} \cdots \sum_{j_n=0}^{d^2-1} c_{j_1 \cdots j_n}^{\sharp} |j_1\rangle_{\sharp} \otimes\cdots\otimes |j_n\rangle_{\sharp}, \notag\\
    c_{j_1 \cdots j_n}^{\sharp} &= \sum_{\alpha_1, \ldots, \alpha_{n-1}=1}^{\chi}  \Gamma_{\sharp\alpha_1}^{[1]j_1}
    \lambda_{\sharp\alpha_1}^{[1]} \Gamma_{\sharp\alpha_1 \alpha_2}^{[2]j_2} \lambda_{\alpha_1}^{[2]}
    \cdots \Gamma_{\sharp\alpha_{n-1}}^{[n]j_n},
\end{align}
where $\{|j_n\rangle_{\sharp}\}_{j=0}^{d^2 - 1}$ is a set of the vectorized matrices $\{ |k_n\rangle\langle l_n| \}_{k,l=0}^{d-1}$. In this case the tensors $\Gamma_{\sharp}^{[k]}$ will each be composed of $\chi d^2$ elements. Consequently the vectorised density matrix in Eq.~\eqref{EqMPSDensity} will possess $nd$ times more parameters than the wavefunction in Eq.~\eqref{EqMPSWavefunc}. It therefore follows that \emph{so long as $2KR < nd$, it will  be more efficient to capture the system behaviour using ERT to construct an ensemble of $R$ MPS wavefunctions rather than evolving a single vectorised density matrix with MPS.} } The reason for this boost is that with the help of principle component analysis ERT finds the optimal basis to represent a given density matrix; whereas, MPS explores the sparsity of the vectorized density matrix.

 
\section{Simulation Results \label{sec:results}}
\label{secComp} 
We now demonstrate the utility of ERT over WMC by applying it to two typical multi-body systems of enduring interest (a Heisenberg spin-chain and a collection of two-level systems coupled to a cavity mode). We choose these systems to be small enough that the Linblad equation can be simulated directly for comparison, but large enough that the ensemble methods are significantly faster. 

We also provide an example of applying ERT to fermionic systems by calculating the power spectrum generated by a driven dissipative Fermi-Hubbard system. The convergence of ERT to the exact result is first checked at a smaller system size, before being applied to a system too large to practically calculate (on the hardware used) the exact dynamics.


In order to assess the performance of the we introduce an integrated error $\mathcal{E}$ for a set of observables $\{\hat{O}_j\}$:
\begin{equation}
    \mathcal{E}=\left(\sum_j\frac{\int {\rm d}t \left(O_j(t)-O_j^A(t)\right)^2}{\int {\rm d}t O^2_j(t)}\right)^{1/2}
\end{equation}
where $O_j(t)$ is the observable expectation from the exact simulation and  $O_j^A(t)$ is the expectation calculated using an approximate method. {In the example simulations considered, when the integrated error is on the order of $10^{-1}$ the accuracy is more than adequate for most purposes, and an error of $10^{-2}$ represents an excellent reproduction of the exact dynamics.}

\subsection{Heisenberg Spin-Chain}

We first consider a simple $N$ site spin-chain system, specifically a Heisenberg XXX model described by the Hamiltonian:
\begin{align}
\hat{H}=-\pi h\sum_{j=1}^{j=N}\hat{\sigma}^{(j)}_z-\pi J\sum_{j=1}^{j=N-1}\bm{\hat{\sigma}}^{(j)} \cdot \hat{\bm{\sigma}}^{(j+1)}.
\end{align}
{ Historically Hamiltonians of this type have been used to model magnetic systems \cite{PhysRevB.98.155102} in order to calculate their critical points and phase transitions \cite{Heisenberg1928}. More recently, this class of models has also been used to study exotic phenomena such as many-body localisation \cite{PhysRevB.77.064426}.}

Given our principal interest in this model is to assess the performance of ERT, in all cases we set $h=J=1$, and initialise the system with all spins in the $+x$ direction (i.e. $\langle \hat{\sigma}^{(j)}_x\rangle=1$). We choose $N=12$ sites corresponding to a Hilbert space dimension of $N_H=2^{12}=4096$. 


\begin{figure}
\centering
\includegraphics[width=\columnwidth]{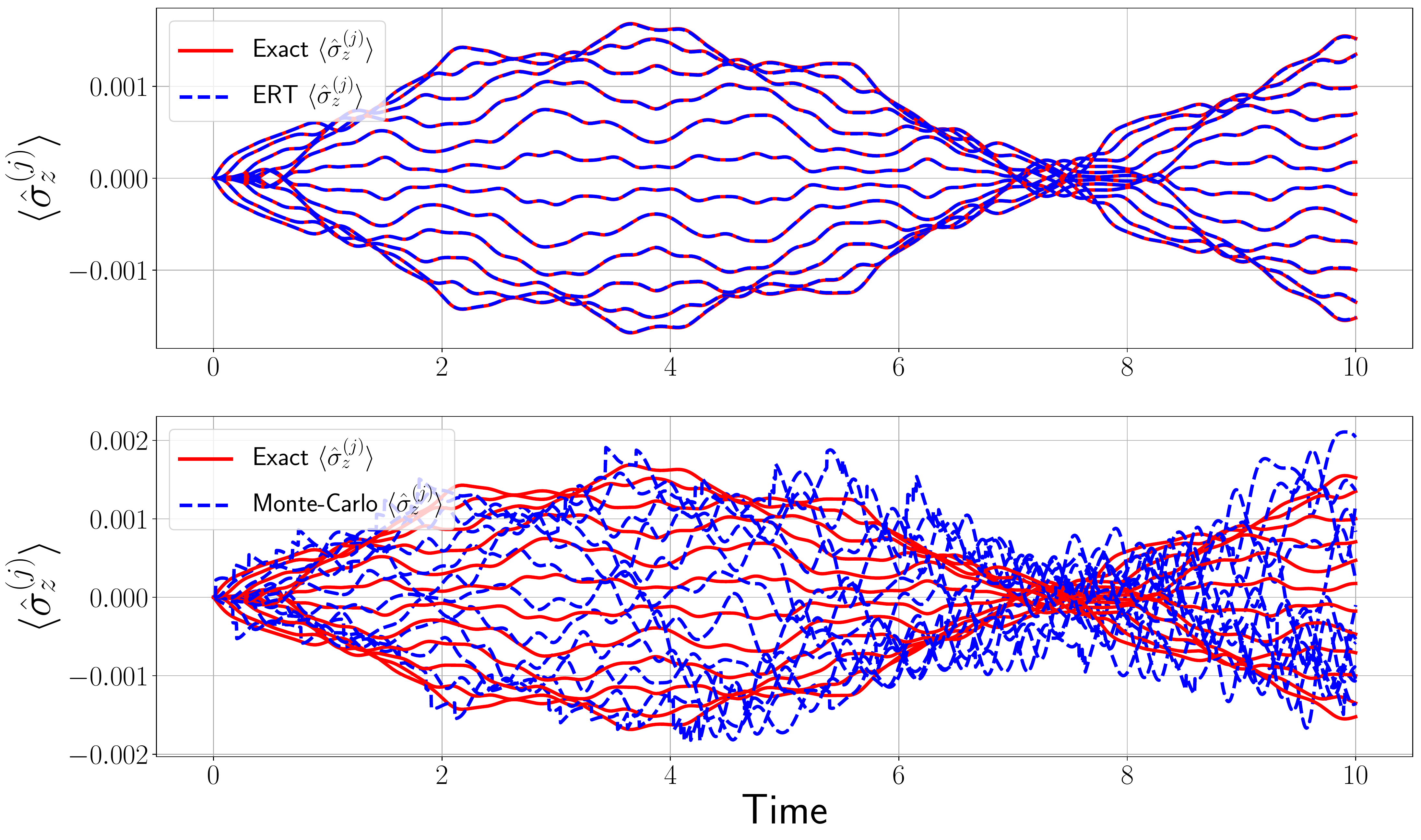}
\caption{Dynamics of $\langle \hat{\sigma}^{(j)}_z\rangle$ (without explicit labelling for individual sites) for $\gamma=0$ and $\Gamma=10^{-3}h$, $\mu=0.9$. The top panel shows the rank 1 approximation, which runs in 159 seconds with an integrated error $\mathcal{E}=1.3\times10^{-2}$. In comparison, a WMC simulation (shown in bottom panel) with 5000 trajectories sampled runs in 4925 seconds with an integrated error of  $\mathcal{E}=3.3$. The exact dynamics runs in 9790 seconds, demonstrating that in this weak environmental coupling regime, ERT significantly outperforms the WMC method in both accuracy and efficiency.}
\label{fig:spinchainexample}
\end{figure}

In addition to the dynamics of the chain itself, we shall include two types of dissipator -- an onsite dephasing
\begin{equation}
    \hat{A}_j=\sqrt{\gamma}\hat{\sigma}_z^{(j)},
\end{equation}
and a set of spin injection/absorption operators at the chain terminals:
\begin{equation}
    \hat{A}^\prime_{1,2}=\sqrt{\Gamma(1\mp\mu)}\hat{\sigma}^{(1)}_\pm, \quad  \hat{A}^\prime_{3,4}=\sqrt{\Gamma(1\pm\mu)}\hat{\sigma}^{(N)}_\pm.
\end{equation}
For both types of dissipator, $\gamma$ and  $\Gamma$ characterise the strength of the environmental coupling to the spin system. For the latter,  $\mu$ biases the driving (analogously to a chemical potential). {Lindblad master equations are only appropriate when the damping is sufficiently slow compared to the internal dynamics. We explore the performance of ERT for damping rates $\Gamma \in [10^{-3}h,10^{-1}h]$, which covers essentially the full range over which Linblad equations are of interest. We explore the behavior both without dephasing and with a weak dephasing rate of $\gamma = 10^{-3}h$.}  

\begin{figure}
    \centering
    \includegraphics[width=\columnwidth]{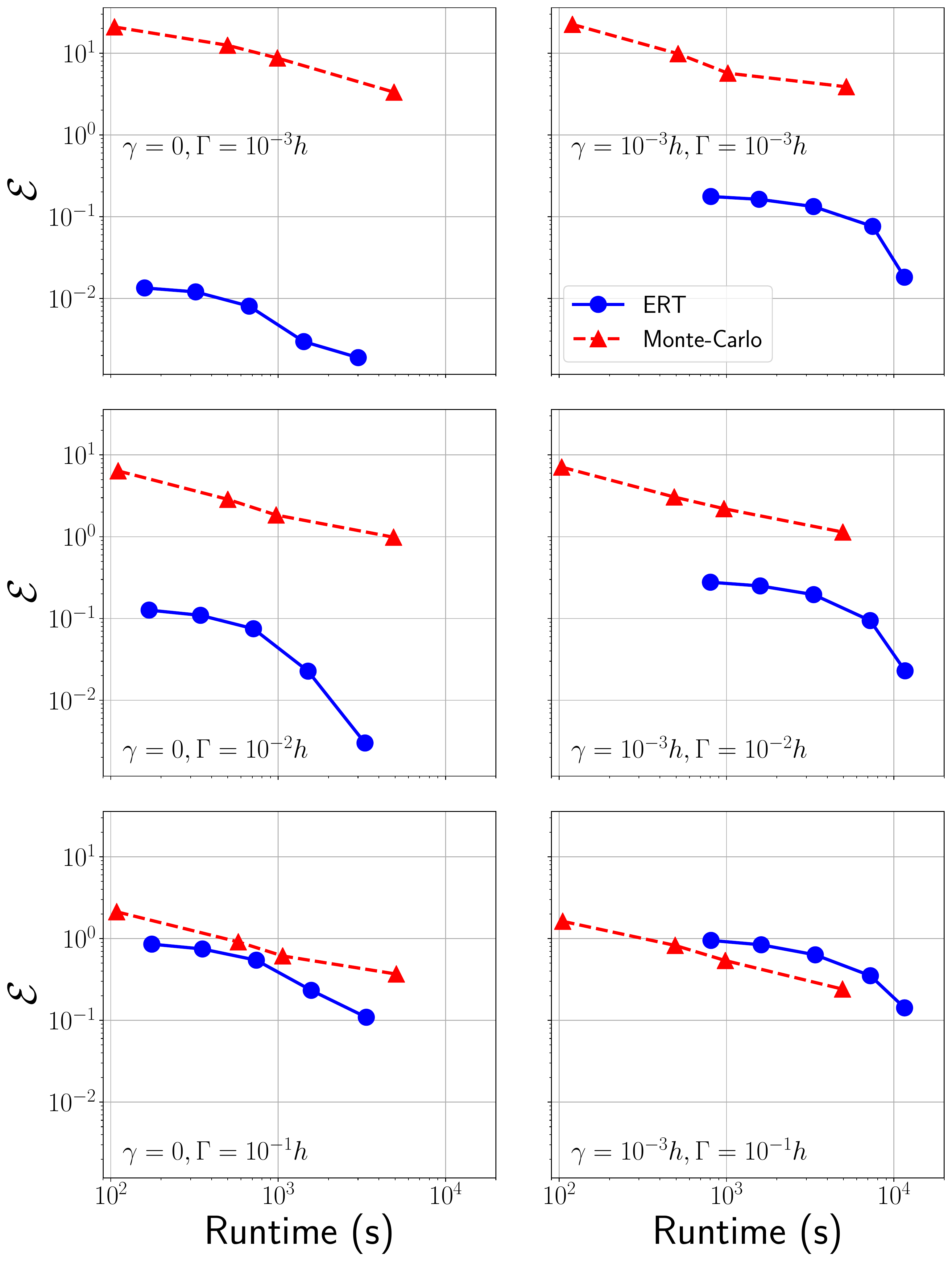}
    \caption{Comparison of spinchain simulation accuracy to runtime for both WMC and ERTs over a variety of dissipative parameters, using $\mu=0.9$. The top panel shows results for $\gamma=0$, while the lower panel uses  $\gamma=10^{-3}h$. Points are generated using the set $R=\{1,2,4,8,12\}$ for ERT, and $n_{traj}=\{100,500,1000,5000\}$ trajectories for the WMC simulations. Exact simulations run in $\sim 9500$ seconds on average.}
    \label{fig:errorsvsruntimesspinchain}
\end{figure}

The numerical simulations of this spin-chain are performed using the Quantum Toolbox in Python (QuTiP)~\cite{JOHANSSON20121760,JOHANSSON20131234}, which includes optimised implementations of both the exact Lindblad master equation, and a parallelized WMC approximation. {All simulations are performed using 4 cores of an Intel Xeon 8280 CPU (with 8 hardware threads), and the ERT method is implemented without any explicit parallelisation. Despite this,  Fig.~\ref{fig:spinchainexample} \& Fig.~\ref{fig:errorsvsruntimesspinchain}, show that \emph{even an unoptimized Python implementation of ERT is able to perfectly reproduce the exact Lindbladian dynamics, improving upon the exact simulation runtime by almost two orders of magnitude. More significantly, the same ERT implementation also significantly outperforms the optimised C implementation of the QuTiP WMC code in both accuracy and runtime.}}  

A more systematic comparison of results is shown in Fig.~\ref{fig:errorsvsruntimesspinchain}, demonstrating the relative performance of the low-rank and WMC methods both with and without dephasing, while varying $\Gamma$ between $\Gamma=10^{-3}h$ and $\Gamma=10^{-1}h$. 
The first conclusion that can be drawn is that ERT yields larger advantages at weaker couplings. This fits with the natural intuition that at weaker couplings the set of wavefunctions generated by the unitaries in Eq.~\eqref{eq:mainformula} will be well characterised by a relatively small set of orthogonal components.  Nevertheless, even at the strongest damping considered, ERT offers comparable accuracy to WMC. In the case of non-zero dephasing, the presence of an additional 12 dissipators reduces the relative accuracy of ERT at lower ranks (although it remains more accurate than the equivalent WMC simulation for the majority of points).  Significantly, runtimes are increased by the presence of additional dissipators, but this is to be expected when the calculation of the new wavefunctions at each timestep is run serially rather than in parallel. In this instance, the runtime for a single timestep will be roughly proportional to $2KR$, as one must apply each of the $2K$ unitaries to the set of $R$ wavefunctions. Since each of these operations is independent however, significant efficiencies could be achieved by parallelising this process when more cores are available.

\subsection{Ensemble of two-level systems in a cavity}
{Our second example is a hybrid system consisting of a single cavity mode coupled to a number of otherwise independent two-level systems (which represent, e.g., cold atoms~\cite{Solano17}, color centers~\cite{Hodges13}, or rare-earth dopants~\cite{Zhong18}). We assume that all the emitters are coupled to the same zero-temperature bath, and the cavity is coupled to a separate input/output port (e.g. a lossy end-mirror). 

In a recent development, a master equation in the Lindblad form has been obtained that accurately describes open systems regardless of the proximity (or otherwise) of the various transitions (and that is valid for all temperatures)~\cite{McCauley2020}. Employing this master equation, placing the two-level systems on resonance with the cavity mode and working in the interaction picture, the effective system Hamiltonian is given by
\begin{align}
   \hat{H} & = \sum_{ij} \hat{\sigma}_+^{(j)} \Lambda_{ij}\hat{\sigma}_-^{(j)}  +  g \sum_j  (\hat{a}^\dagger \hat{\sigma}_-^{(j)}  + a \hat{\sigma}_+^{(j)} ) \\  
   & \;\;\; + \sqrt{\kappa}\beta (\hat{a} + \hat{a}^\dagger). 
\end{align} 
Here $\Lambda_{ij} = \Delta_i \delta_{ij} + \sqrt{\lambda_i \lambda_j}$ includes the detuning between emitter $i$ and the cavity, $\Delta_i$, as well as the respective bath-induced Lamb shifts, $\lambda_i$. Here $\hat{\sigma}_-^{(j)}$ is the lowering operator for the $j$th atom from its excited to its ground state and $\hat{a}$ is the annihilation for the cavity mode. The term proportional to $g$ is the coupling between the emitters and the cavity mode, and the last term describes driving of the cavity with a coherent state at the rate of $|\beta|^2$ photons per second~\cite{Wiseman2010-hl}. The Lindblad dissipators for the system are given by 
\begin{equation}
    \hat{A}_{1}=\sqrt{\kappa}\hat{a}, \quad  \hat{A}_{2}=\sqrt{\gamma}\sum_j\hat{\sigma}_-^{(j)} , 
\end{equation}
which describes output coupling of the cavity mode at rate $\kappa$ and collective decay of the exited states of all the emitters at the rate $\gamma$. } 

\begin{figure}
\centering
\includegraphics[width=\columnwidth]{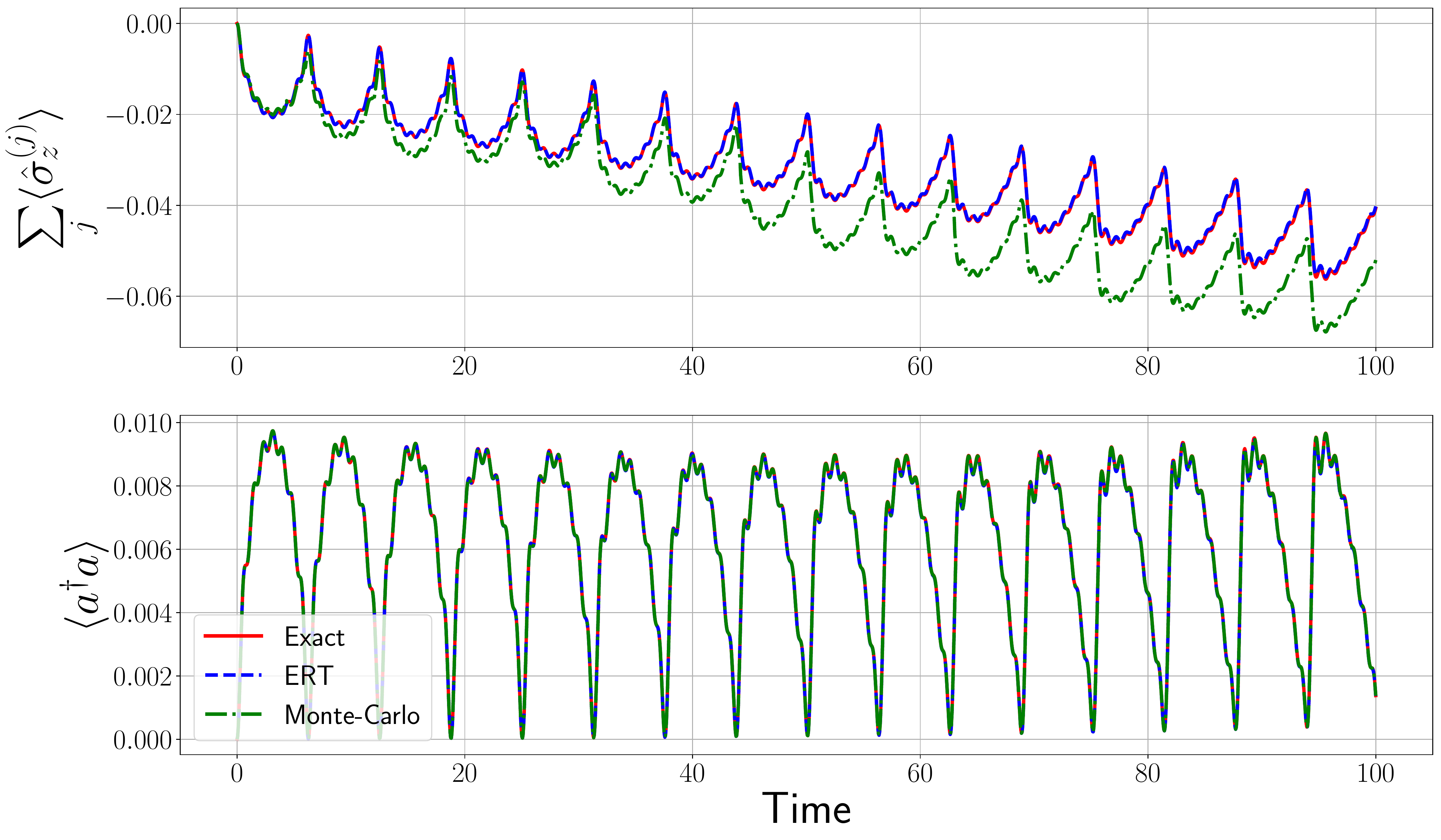}
\caption{Dynamics of the atomic ensemble in a cavity for $\gamma=10^{-3}g$ and $\beta=\kappa=10^{-1}g$, with the exact, rank 1 ERT, and $5000$ trajectory WMC dynamics shown. The top panel shows the collective atomic dynamics while the bottom panel demonstrates the cavity occupation numbers. The exact simulation takes 5975 seconds to compute, while the WMC simulation runs in 736 seconds, with an integrated error over all expectations of $\mathcal{E}=1.0$. In comparison the ERT computation completes in 32 seconds with an integrated error of $\mathcal{E}=2.0\times10^{-2}$, improving on WMC by almost two orders of magnitude in both accuracy and runtime.}
\label{fig:cavityexample}
\end{figure}

\begin{figure}
    \centering
    \includegraphics[width=\columnwidth]{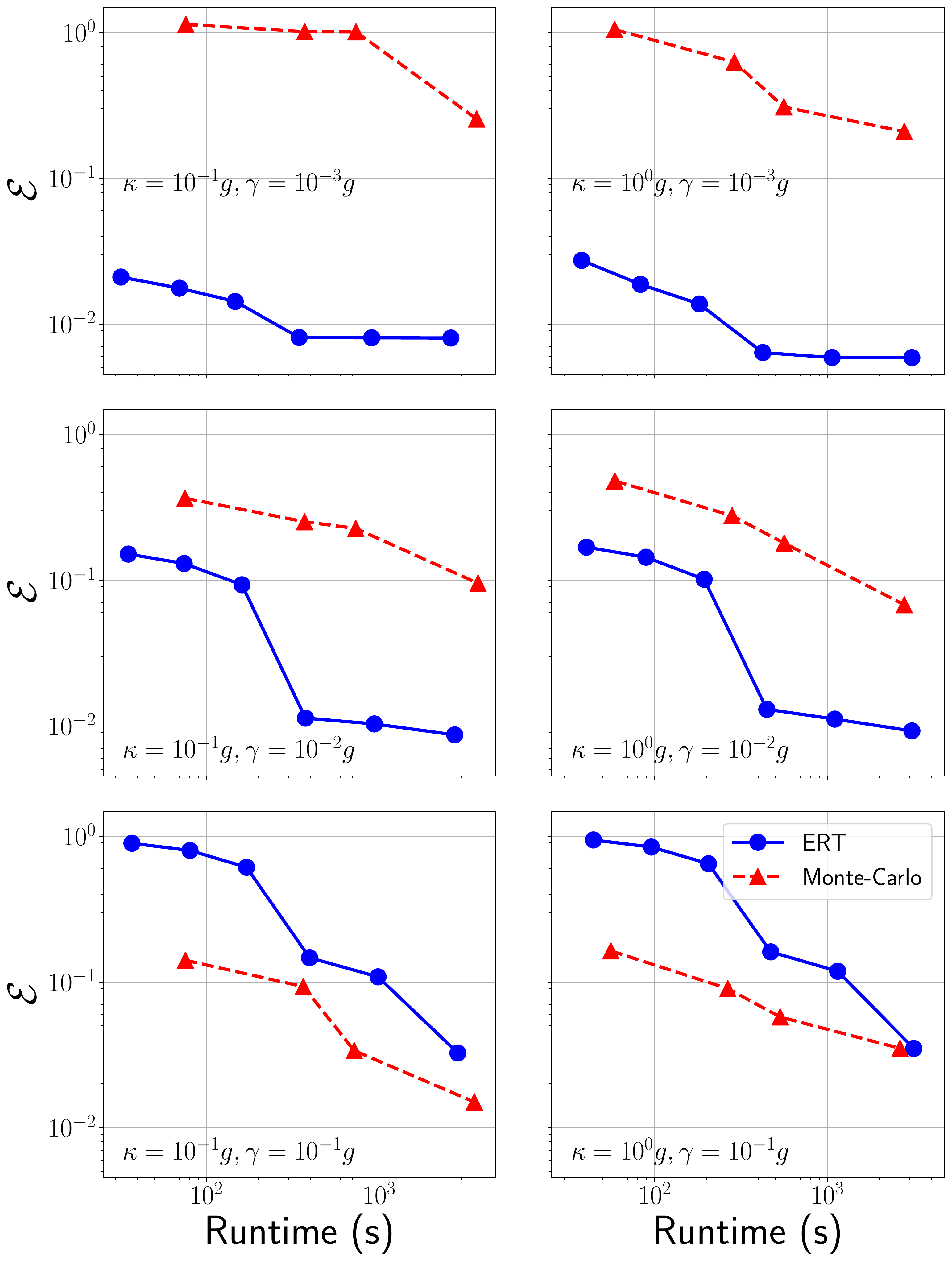}
    \caption{Comparison of spinchain simulation accuracy to runtime for both WMC and ERTs over a variety of dissipative parameters. Once again we see that ERT provides a decisive advantage at weak couplings, but requires higher ranks to properly capture dynamics at stronger couplings. The average exact master equation simulation time is shown as a dashed vertical line on both graphs. Points are generated using the set $R=\{1,2,4,8,16,32\}$ for ERT, and $n_{traj}=\{100,500,1000,5000\}$ trajectories for the WMC simulations. Exact simulations run in $\sim 5500$ seconds on average.}
    \label{fig:cavityerrors}
\end{figure}

Once again we focus on the regime of weak coupling, setting $g=1$, $\Lambda_{ij}=20jg \delta_{ij}$ and $\kappa=\beta$ For all simulations we take 8 atoms and initialise each atom in an equal superposition of its ground and excited states, along with an empty cavity. Simulations are run with the same resources as in the previous section, and as Fig.~\ref{fig:cavityexample} demonstrates, ERT is able to outperform WMC in capturing the decay of the atomic ensemble at a smaller computational cost. When varying $\gamma$ and $\kappa$, Fig.~\ref{fig:cavityerrors}  shows once again that for weak couplings ERT provides a clear advantage over WMC in both efficiency and accuracy. Even at the largest $\gamma$ considered, we see that at higher ranks the performance of ERT becomes comparable to WMC. This trend is repeated both when including off-diagonal elements in  $\Lambda_{ij}$, and initialising the system from different states (not shown). 


\subsection{Dissipative Fermi-Hubbard model} 
\label{secFH}

Finally we apply the ERT method to a fermionic system, specifically the Fermi-Hubbard model ~\citep{Tasaki1998}. This model is a useful benchmark, given its status as one of the paradigmatic models for strongly correlated electronic systems. Its Hamiltonian is given by
\begin{align}
\hat{H}\text{\ensuremath{\left(t\right)}}= & -t_{0}\sum_{j\sigma}\text{\ensuremath{\left( \hat{c}_{j\sigma}^{\dagger}\hat{c}_{j+1\sigma}+\hat{c}_{j+1\sigma}^{\dagger}\hat{c}_{j\sigma}\right)}}\nonumber \\
 & +U\sum_{j}\hat{c}_{j\uparrow}^{\dagger}\hat{c}_{j\uparrow}\hat{c}_{j\downarrow}^{\dagger}\hat{c}_{j\downarrow},
\label{eq:Hamiltonian}\end{align}
where $\hat{c}_{j\sigma}$ is the fermionic annhilation operator for the $j$th site and spin $\sigma$, $t_0$ is the hopping parameter and $U$ is the onsite potential. The Fermi-Hubbard model is of particular interest as under coherent external driving it exhibits a high degree of complexity in its dynamics ~\cite{mccaul2020controlling,mccaul2020driven}. When dissipation is added this model contains a number of intriguing features, including symmetry breaking phase transitions ~\cite{Sponselee2019} and dynamic reversals of magnetic correlations ~\cite{PhysRevLett.124.147203}. 

To incorporate dissipation, we apply fermion injection/absorption operators for each species to the terminals of an $N$ site chain
\begin{align}
    \hat{A}^\prime_{1,2}&=\sqrt{\Gamma(1-\mu)}\hat{c}_{1\{\uparrow,\downarrow\}}, \quad  \hat{A}^\prime_{3,4}&=\sqrt{\Gamma(1+\mu)}\hat{c}^\dagger_{1\{\uparrow,\downarrow\}} \notag \\
      \hat{A}^\prime_{5,6}&=\sqrt{\Gamma(1+\mu)}\hat{c}_{N\{\uparrow,\downarrow\}}, \quad  \hat{A}^\prime_{7,8}&=\sqrt{\Gamma(1-\mu)}\hat{c}^\dagger_{N\{\uparrow,\downarrow\}} 
\end{align}
and use these dissipators to drive the system from its (isolated) half-filled ground-state.
\begin{figure}
    \centering
    \includegraphics[width=\columnwidth]{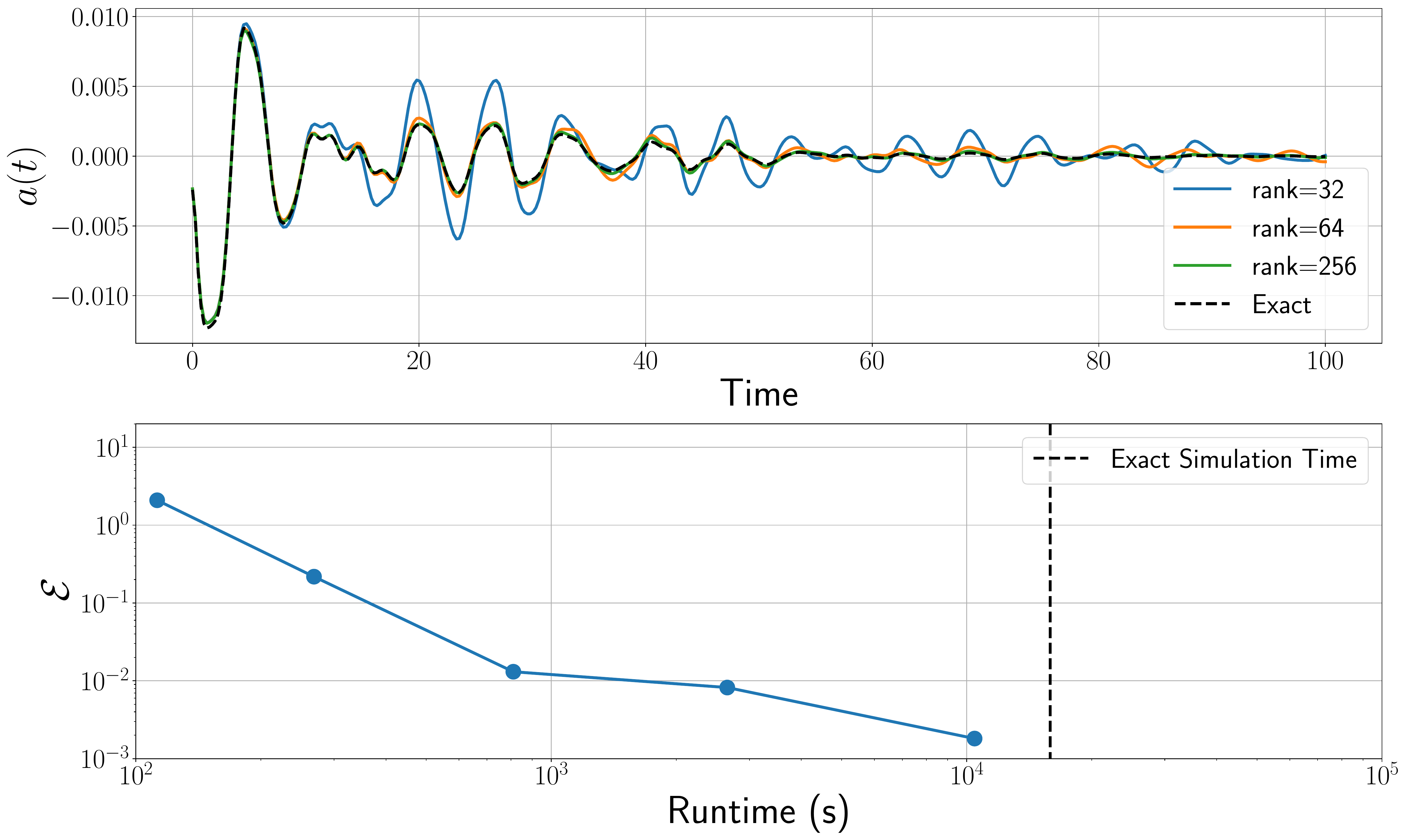}
    \caption{Calculation of the dipole acceleration $a(t)=\frac{{\rm d} J(t)}{{\rm d}t}$  in a $U=t_0$, $N=6$ site system at a relatively strong coupling $\Gamma= 0.03 t_0$ and bias $\mu=0.9$. The top panel shows the time dynamics of $a(t)$, while the bottom panel demonstrates the relative accuracy and efficiency of the approximation using $R=\{16,32,64,128,256\}$.}
    \label{fig:FermiHubbardFirstTest}
\end{figure}

In order to simulate this model, we use the Python package QuSpin ~\cite{10.21468/SciPostPhys.7.2.020,SciPostPhys.2.1.003}, which is capable of efficiently simulating fermionic systems. Note that as QuSpin lacks an efficient WMC solver for dissipative systems, in this example we forgo comparison to approximate methods.

In the case of a coherently driven, non-dissipative system, the electronic current 
\begin{equation}
\hat{J}\text{\ensuremath{\left(t\right)}}=  -t_{0}\sum_{j\sigma}\text{\ensuremath{\left( \hat{c}_{j\sigma}^{\dagger}\hat{c}_{j+1\sigma}-\hat{c}_{j+1\sigma}^{\dagger}\hat{c}_{j\sigma}\right)}}.
\end{equation}
may exhibit an optical phenomenon known as High Harmonic Generation (HHG). HHG occurs when the dipole acceleration $a(t)=\frac{{\rm d} J(t)}{{\rm d}t}$ spectrum has a  a highly non-linear response to driving, generating frequencies many multiples larger than the driving field ~\citep{Ghimire2012, Ghimire2011a, Murakami2018}. It is therefore natural to ask if similar non-linearities are present when one uses incoherent dissipative driving. 
As an initial test of the applicability of ERT to this model, we calculate the dipole acceleration $a(t)$ in a $U=t_0$, $N=6$ ($N_H=4^6=4096$) system and compare it to the results of an exact simulation.  As Fig.~\ref{fig:FermiHubbardFirstTest} shows, once again ERT is able to obtain good accuracy while still significantly improving on the time taken to run the calculation via exact methods.

\begin{figure}
    \centering
    \includegraphics[width=1\columnwidth]{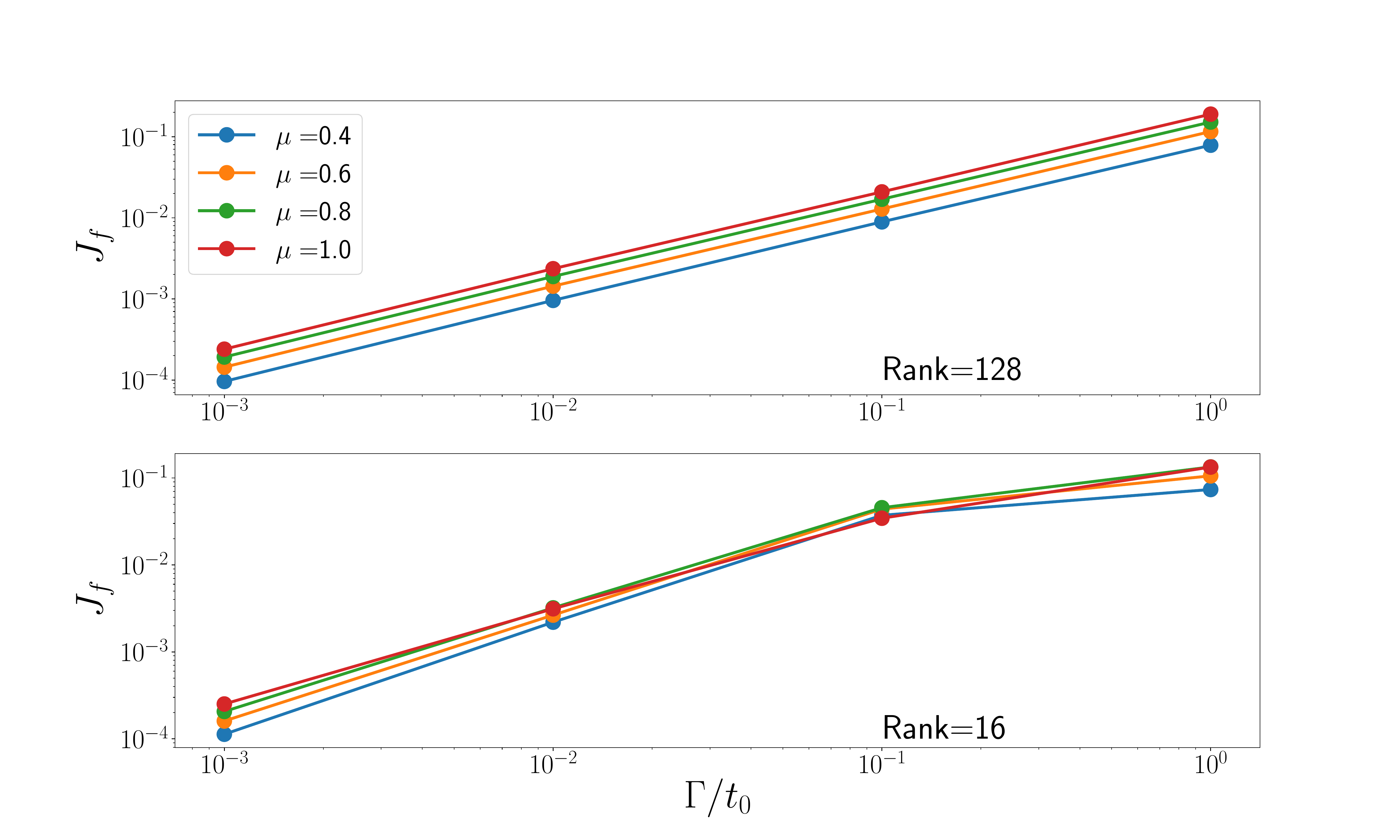}
    \caption{Plotting steady state currents $J_f$ against coupling strength. One finds that at higher ranks the expected linear relationship is recovered, while lower ranks do not capture the final state sufficiently well to maintain this linearity. Note however that at weaker couplings results agree reasonably well between both $R=16$ and $R=128$, and it is only at stronger couplings that one is required to use a larger rank to obtain accurate results.}
    \label{fig:SteadyState}
\end{figure}

Using ERT we are able to investigate whether or not non-linear effects are present in this dissipatively driven model at larger system sizes than can be practically calculated using exact methods. As an example, we consider the $U=0$, $N=14$ Fermi-Hubbard model with dissipative driving. In this free case, Hamiltonian symmetries reduce the effective size of the Hilbert space to $N_H=16384$, but this is still well beyond the dimension at which an exact master equation calculation is practical without significant computational resources. One can test the accuracy of the approximation in this scenario in a number of ways -- the most straightforward being to check that as the rank $R$ is increased, simulation results converge. A second check in this instance is utilising the fact that the Hubbard model is known to exhibit diffusive transport when dissipatively driven ~\cite{PhysRevB.86.125118}. From this, one would expect that the steady state currents $J_f$ should depend linearly on the coupling strength. Figure~\ref{fig:SteadyState} shows that at sufficiently large ranks this is the case, providing a further check for the accuracy of ERT. 

\begin{figure}
    \centering
    \includegraphics[width=1.1\columnwidth]{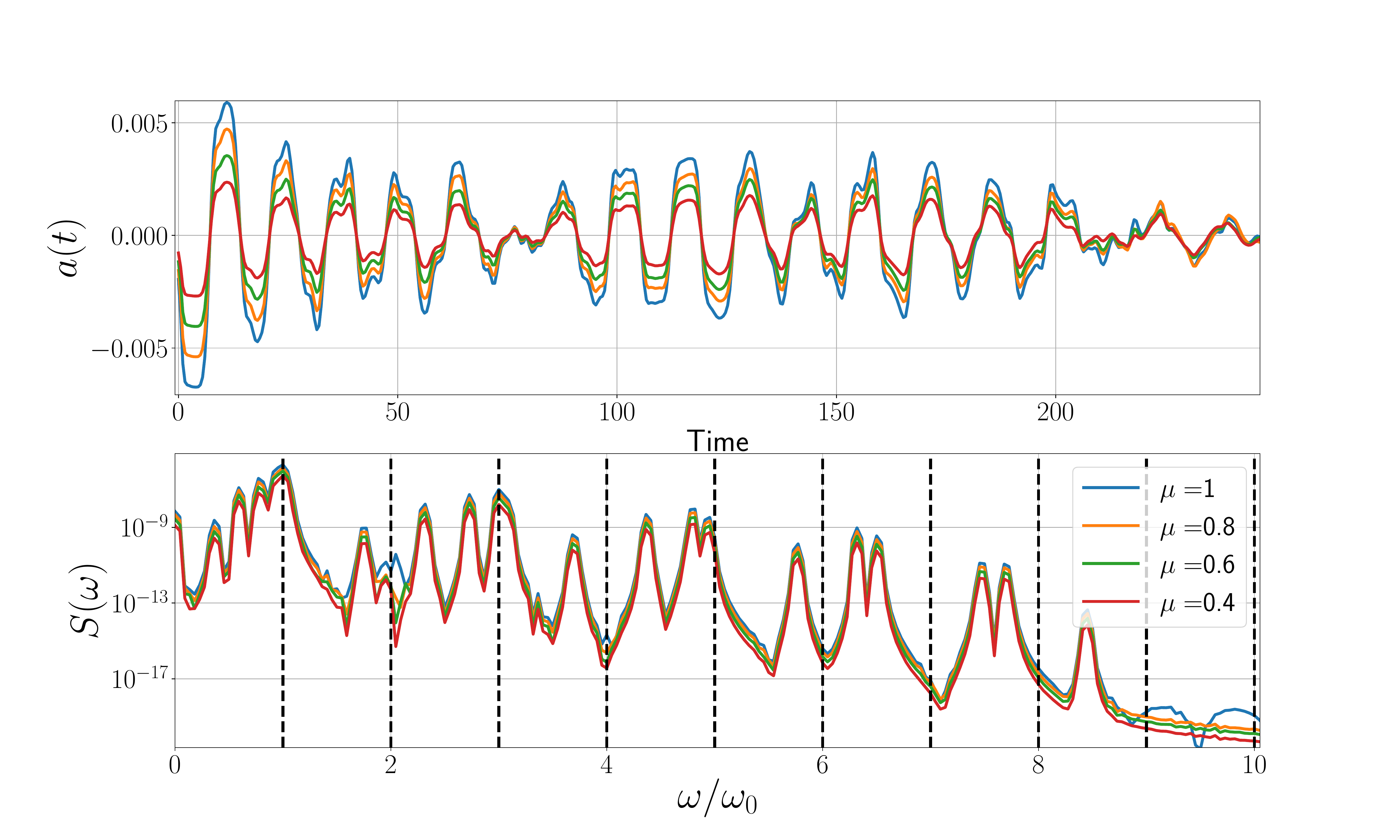}
    \caption{Dipole acceleration and its frequency spectrum for a $U=0$, $N=14$ site system with a relatively strong coupling $\Gamma= 0.1 t_0$. One finds good convergence at $R=128$, with the spectrum displaying a number of prominent frequencies. }
    \label{fig:FermiHubbardHHG}
\end{figure}

Figure~\ref{fig:FermiHubbardHHG} demonstrates that incoherent dissipative driving generates non-linear effects in the transient dipole acceleration, analogously to the coherent external driving which produces HHG. In this instance, without an external driving frequency to compare to, we instead define a `fundamental frequency'  $\omega_{0}$ as the most strongly represented frequency by taking $S(\omega)=\left|  FFT\left\{a(t)\right\} \right|^2$ and setting
\begin{equation}
    \max\limits_\omega S(\omega)=S(\omega_{0}).
\end{equation}

 Examining the dipole acceleration spectrum, we find that the transient current contains number of distinct frequencies. This is particularly interesting when contrasted to the case of an isolated $U=0$ system driven by an external transform-limited pulse, where overtones only appear at odd integer multiples of the driving frequency  ~\cite{Silva2018}. In the present case,  while we observe a prominent peak in the spectrum at the third harmonic, there are also a number of off-integer and sub-linear frequencies present. The position of these peaks (including $\omega_0$) remain constant over a wide range of $\Gamma$ and $\mu$, indicating that while the transient system response to constant incoherent driving is non-linear, the dissipative parameters meaningfully affect only the size of that response rather than its spectral character. 
 
\section{Discussion \label{sec:discussion}}
We have introduced an approximate method for the fast simulation of dissipative quantum systems. This technique is based upon first representing an infinitesimal Lindblad evolution in a Kraus form, from which one can construct an ensemble of wavefunctions which capture the full dynamics of the system. When the size of the ensemble exceeds a prespecified rank $R$, it is truncated in the space of its principal components. We have termed this method Ensemble Rank Truncation (ERT), and have found a number of significant results. First, this method may employed in combination with Matrix Product State (MPS) wavefunction decompositions, and under certain easily satisfiable conditions ERT + MPS will in principle be more efficient than either MPS alone or MPS + WMC in simulating large dissipative systems. 

The performance of ERT was investigated in a number of physically distinct systems, allowing for direct comparison to  WMC simulations. Here it was found that in the regime of weak dissipative coupling, ERT offered order of magnitude improvements in both accuracy and computing time compared to WMC. {This result is at least partially due to the difference between ERT and WMC in their rate of convergence to the true solution. In the latter case, WMC asymptotically converges at $O(\sqrt{n_{\rm traj}})$, while the convergence of ERT with $R$ will depend upon the specifics of the system dynamics. While one cannot determine \emph{a priori} the appropriate rank to use with ERT, it should be possible to use system identification techniques \cite{Lee2013,VanWingerden2007,Juang2005} to ascertain the effective rank of the system before beginning an ERT simulation. As we have seen in all the cases considered, the rank $R$ required to accurately model a system with ERT is much lower than either the Hilbert space dimension $N_H$ or the number of trajectories $n_{\rm traj}$ required by WMC.

An important question is how ERT will scale in parallel implementations using many processors. For WMC each trajectory is independent, and the speedup from multiprocessing will be approximately linear \cite{JOHANSSON20121760}. For ERT, each infinitesimal Kraus map may also be independently applied to each element of the ensemble. This part of the scheme is therefore trivially parallelisable and we would expect a similar linear scaling. Although the orthogonalisation step requires some cross-communication to calculate the overlap matrix $S$, we found in Sec. \ref{sec:Derivation} that the cost of this step compared to evolving an individual element of the ensemble will be small when $N_H \gg L^2$. We therefore expect that in this scenario, a properly parallelised ERT implementation will scale with number of processors approximately as well as WMC.}

While the advantages of ERT have already been demonstrated, there remains a good deal of scope for improving the method. For instance, one might choose to adopt an adaptive rank $R$ at each step. As a simple example, one could improve accuracy by choosing $R$ at each step such that the sum of discarded eigenvalues in Eq.~\eqref{eq:diageigenvalues} is below some tolerance. Furthermore, it should be possible to improve on the $O({\rm d}t^2)$ error found in Eq.~\eqref{eq:mainformula}. This could be achieved either by re-deriving $\hat{\mathcal{U}}_k, \hat{\mathcal{V}}_k$ using higher order Suzuki-Trotter decompositions ~\cite{Hatano2005}, or through the application of Richardson extrapolation to Eq.\eqref{eq:mainformula}. A final avenue for extension would be to improve the domain of applicability of the ERT, deriving the infinitesimal Kraus operators for the non-Markovian generalisation of the Lindblad equation ~\cite{PhysRevA.75.022103}.

In summary, ERT is a novel technique for the simulation of dissipative quantum systems that is capable of outperforming existing techniques, and can in principle be alloyed to state-of-the-art methods for approximating large quantum systems.  

\end{document}